\documentclass[twocolumn]{article}
\usepackage{graphicx}
\begin{document}
\title{Weak response of nuclei}
\author{A. Botrugno and G. Co' \\
Dip. Fisica Universit\`a di Lecce and
          INFN sez. di Lecce, I-73100 Lecce, ITALY}
\date{}
\maketitle
{\small ABSTRACT:
  We discuss some differences and similarities between
  electron and neutrino scattering off atomic nuclei.  We find that,
  in the giant resonance region, the two processes excite different
  nuclear modes, therefore the weak and the electromagnetic nuclear
  responses are rather different.  In any case, the scattering of
  electrons and photons is the best guide we have to test the validity
  of our nuclear models and their prediction power.  The experience in
  describing electromagnetic excitations of the nucleus, suggests
  that, when the nucleus is excited in the continuum, the
  re-interaction between the emitted nucleon and the remaining nucleus
  should not be neglected.  A simple model taking into account this
  final state interaction is proposed, and applied to the neutrino
  scattering off $^{16}$O nucleus.}

\vskip 1.0 cm
The great activity of the last fifteen years in neutrino physics has
attracted great attention to the interaction between neutrinos and
atomic nuclei.  At present, the main interest in the neutrino-nucleus
interaction is related to the goal of investigating the properties of
the neutrinos, or of the neutrino sources such as stars, supernovae,
 earth etc. .  In this perspective the nucleus is
considered as detector, and therefore the nuclear response to weakly
interacting probes should be well controlled.

Our knowledge about the interaction of electrons and photons with the
nucleus can be used as a guide to make prediction about
neutrino-nucleus processes. Both electromagnetic and weak interactions
can be well described within a perturbation expansion of the
scattering amplitudes.  Furthermore, the tensor structure of the
electromagnetic current is identical to that of the vector part of the
weak current.
\begin{figure}[ht]
\resizebox{0.5\textwidth}{!}
{\includegraphics[angle=0]{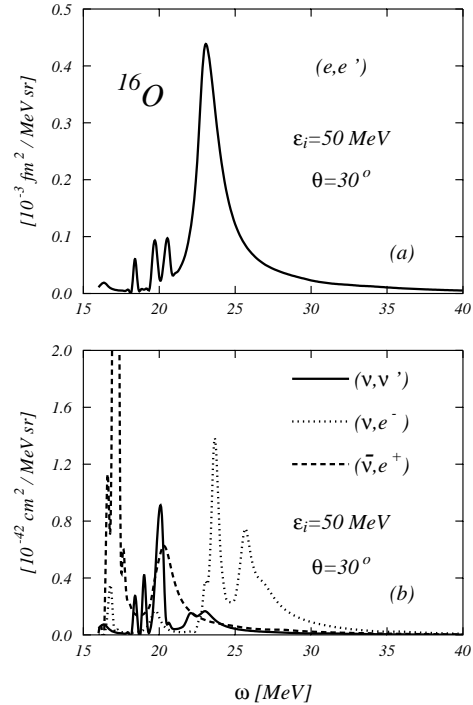} }
\caption{Doubly differential
  cross sections $d^2 \sigma / d \epsilon_f d \Omega$ 
  for scattering of electrons, panel 
  (a), and neutrinos, panel (b), as a function of the nuclear
  excitation energy. The lepton incoming energy $\epsilon_i$, and the
  scattering angle are the same for all the reactions considered.}
\label{fig:comp}
\end{figure}

In fig.\ref{fig:comp} we present a direct comparison between electron
and neutrino double differential cross sections off $^{16}$O target
nucleus.  The results, shown in the figure as a function of the
nuclear excitation energy, $\omega$, have been obtained for the same
values of the projectile energy $\epsilon_i$= 50 MeV, and of the
scattering angle $\theta$ = 30$^o$.  The cross sections have been
calculated in first order plane wave Born approximation, i.e.
considering the exchange of a single boson, either a photon, a $Z^o$
or a $W^\pm$, and by describing the lepton wave functions in terms of
plane waves.  The nucleus is excited above the nucleon emission
threshold, and we describe the transition of the nucleus from its
ground state to the excited state in the continuum region by using the
Continuum Random Phase Approximation (CRPA) whose equations are solved
as described in \cite{deh82,co85}.  The results shown in the figure
have been obtained by using a zero range Landau-Migdal interaction.
The cross sections have been obtained by summing all the positive and
negative multipole excitations up to a maximum value of the total
angular momentum, $J=6$.
\begin{figure}
\resizebox{0.5\textwidth}{!}
{\includegraphics[angle=0]{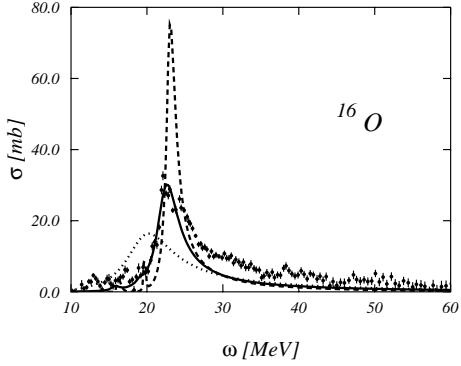} }
\caption{Total photoabsorption cross section compared with the data of
ref. \cite{ahr75}. The dashed lines shows the continuum RPA results,
the dotted lines have been obtained using the folding with asymptotic
parameters, the full lines using the energy dependent folding.}
\label{fig:phot}
\end{figure}
\begin{figure}[ht]
\begin{center}
\includegraphics[angle=0,scale=0.6]{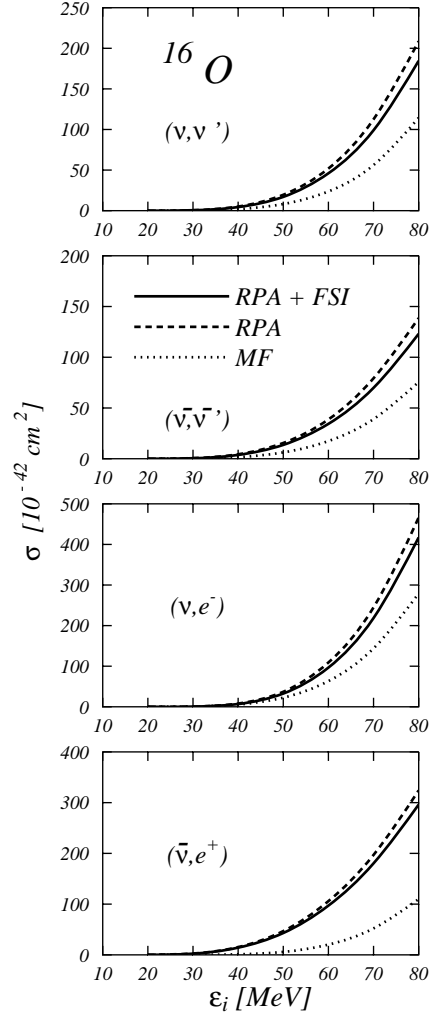}
\end{center}
\vskip -1.0 cm 
\caption{Total inelastic cross sections as a function of the neutrino,
or antineutrino, incoming energy.}
\label{fig:tot}       
\end{figure}

The shapes of the cross sections shown in fig.\ref{fig:comp} are
rather different. The difference between electron and charge exchange
neutrino processes was expected, since the basic particle-hole
transitions inducing charge exchange excitations are quite different
from those involved in the charge conserving excitations.  The
noticeable difference between electron and charge conserving neutrino
scattering is more surprising, since in this case the basic
particle-hole transitions are the same. We made a multipole
decomposition of the cross sections to understand this difference, and
we found that the relevant multipoles forming the cross sections are
different for the electron and neutrinos.  The $1^-$ excitation is
responsible for the 93\% of the electron scattering cross section.
The remaining is due for the 6\% to the $2^+$ and for about a 1\% to
the $0^+$. The situation is quite different for the ($\nu,\nu'$) cross
section: 58\% $2^-$, 33\% $1^-$, 6\% $0^-$, 2\% $1^+$ and about 1\%
$3^+$.
This difference between electron and neutrino scattering is due to the
fact that in neutrino scattering the nuclear transitions are dominated
by the axial transverse operator, absent in electromagnetic
excitations.  This fact is not related to the specific kinematics of
fig. \ref{fig:comp} but it is more general.  We made calculations also
in the quasi-elastic regime, with $\epsilon_i$=1 GeV. In this case the
shapes of the various cross sections are very similar, showing a
single large peak at the same value of the nuclear excitation energy.
However, even in this case, for the neutrino scattering the main
contribution to the cross section is induced by the axial current
operator.  These results suggest caution in the comparison between
electron and neutrino cross sections. While in electromagnetic
excitations the vector transition operator excites mainly natural
parity states, the nuclear excitations produced by neutrinos are ruled
by the axial vector part of the transition operator which excites
without any preference both natural and unnatural parity states.  For
this reason, being able to reproduce electron scattering cross
sections does not necessarily imply a good control of the neutrino
cross sections.

In spite of the words of caution expressed above, electromagnetic
excitations are still the best benchmarks we have to test our
description of nuclear excitations.  The limits of our capacity of
describing nuclear excitations in the continuum region can be well
summarized by the results shown in fig. \ref{fig:phot} where we
compare the experimental total photon absorption cross sections of the
$^{16}$O nucleus \cite{ahr75} with various theoretical cross sections.
The dashed line show the result obtained with the CRPA used in fig.
\ref{fig:comp} \cite{deh82,co85}. The features of these calculations
are well known in the literature and they are common to all the
continuum RPA results.  They are rather independent from the residual
interaction and from the technique used to treat the continuum. While
the position of the resonance is well reproduced, the CRPA cross
sections overestimate the size of the experimental cross sections and
they underestimate their width.  It is commonly believed that these
problems could be solved by considering many-body effects beyond RPA,
such as many-particle many-hole excitations \cite{dro90}.

Many body effects are relevant not only in the giant resonance region
but also in the quasi-elastic peak.  Also in this case the CRPA
results do not provide a good description of experimental data
\cite{co88}.  The inclusion of many body effects beyond RPA, called in
this context final state interactions (FSI), greatly improves the
agreement with the data \cite{co88,ama94,fab89}.  To take into account
the effect of the FSI in the quasi-elastic region we have developed a
phenomenological model \cite{co88}. After assuming that the FSI are
independent from the multipolarity of the nuclear excitation we can
express the FSI response $S^{FSI}$ in terms of the RPA response
$S^{RPA}$ as:
\begin{eqnarray}
&~&
\nonumber
S^{FSI}(|{\bf q}| , \omega)  = \\
\nonumber
&~&
\int_0^\infty dE \,\,\,
S^{RPA}(|{\bf q}| , E) 
\left[ \rho(E,\omega)+\rho(E,-\omega) \right]
\end{eqnarray}
where the folding function is given by:
\[
\rho(E,\omega)=\frac{1}{2\pi}
\frac{\Gamma(\omega)}
{ 
[ E-\omega-\Delta(\omega) ]^2 +
[ \Gamma(\omega)/2]^2 
}
\]
The quantities $\Delta$ and $\Gamma$ are linked by a dispersion relation
\[
\Delta(\omega)= \frac{1}{2 \pi} P
\int_0^\infty d\omega' \,\, \frac{\Gamma(\omega')}
{\omega'-\omega}
\]
therefore we only have to fix the values of the $\Gamma(\omega)$
function. We used the prescription of taking the energy average
between the single particle widths of both particle and hole wave
functions:
\[
\Gamma(\omega) \sim  
\frac{1}{\omega} \int_0^\omega d \epsilon
 \left[ \gamma(\epsilon_F + \epsilon +\omega) + 
        \gamma(\epsilon_F  + \epsilon - \omega )
\right]
\]
The $\gamma$ widths are fixed to reproduce the values of the volume
integrals of the imaginary part of the optical potential \cite{mah81}. 
We have used the following parameterization:
\begin{equation}
\gamma(\epsilon)= A_\Gamma 
\left( \frac {\epsilon^2} {\epsilon^2 + B_\Gamma^2 } \right)
\left( \frac {C_\Gamma^2} {\epsilon^2 + C_\Gamma^2 } 
\right)
\label{eq:par}
\end{equation}
with $ A_\Gamma $=11 MeV, $ B_\Gamma $=20 MeV and $ C_\Gamma $=110
MeV. Our CRPA results, corrected in this way to consider the FSI
effects, reproduce rather well the quasi-elastic longitudinal and
transverse responses of $^{12}$C and $^{40}$Ca \cite{ama94} and the
total inclusive cross section of $^{16}$O \cite{co02}.

The success of this model in reproducing the quasi-elastic responses
pushed us to adopt it also in the giant resonance region.  The result
obtained by a straightforward application of the model to the total
photoabsorption cross section is shown by the dotted line of fig.
\ref{fig:phot}. Evidently the FSI effects are overestimated.  The
values of the constants have been fixed to reproduce many body effects
which modify mainly the motion of the single particle as if the
nucleon moves in an optical potential.  These are the most important
effects in the quasi-elastic region, but in the giant resonance region
more complicated many-particle many-hole excitations become important
as it has been shown in \cite{dro87}.  The approximation of using
optical potential parameter is also related to the assumption of the
independence of the FSI from the multipole excitation.  This
assumption is plausible in the quasi-elastic regime where many
multipoles contribute to the total cross section with comparable
strength, but it is hardly justified in the giant resonance regime,
usually dominated by few excitation multipoles.

To take into account the peculiarities of the giant resonance region
we modify the values of the parameters of eq. \ref{eq:par}. We used a
set of energy dependent parameters. For energies above the giant
resonance region ($\omega >$ 40 MeV) we use the asymptotic values
given above.  We fix for $\omega$ = 10 MeV the values $ A_\Gamma $ =
6 MeV and $ B_\Gamma $ = 60 MeV and we let them evolve linearly as a
function of the excitation energy $\omega$ up to their asymptotic
values. These parameters have been fixed to reproduce at best the
photoabsorption data. The resulting cross section is shown in
fig. \ref{fig:phot} by the full line.

The procedure described above has been used also for the $^{12}$C
nucleus, where we could test the validity of our model by comparing
its results with the (few) inclusive electron scattering data in the
giant resonance region \cite{bar83}. Also in this case the inclusion of
FSI improves noticeably the agreement with the data.
\begin{figure}
\resizebox{0.5\textwidth}{!}
{\includegraphics[angle=0]{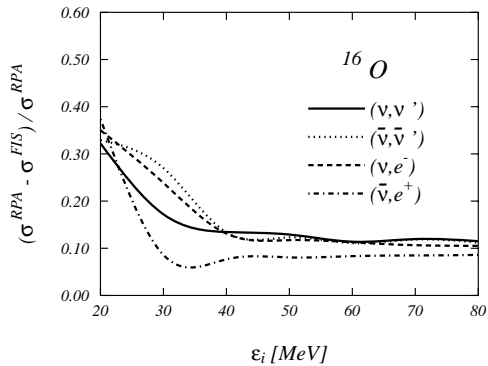} }
\caption{Relative differences between the cross sections of fig. \ref{fig:tot}}
\label{fig:rat}
\end{figure}

The results of fig. \ref{fig:phot} show that the main effect of the
FSI is to move strength from the peak of the resonance to higher
energies.  For neutrinos of few tens of MeV part of strength goes in a
region kinematically forbidden. This implies that the total cross
section is reduced with respect to that predicted by the CRPA. This
effect is evident in fig. \ref{fig:tot} where we show the total
inclusive cross sections for various neutrino and antineutrino
reactions as a function of the neutrino incoming energy. The CRPA
results are shown by the dashed lines, while the inclusion of the FSI
effects produces the full lines. For sake of comparison also the
results obtained with a pure mean field model are shown (dotted
lines).

The first, rather obvious remark, is the large difference between mean
field and CRPA prediction, showing the inadequacy of the mean field in
describing the cross sections in this energy region.  The second
remark is that, as we have expected, the FSI reduce the total cross
section.  This effect is more relevant at low neutrino energies, since
for an analogous shift of the strength, the kinematically forbidden
region, is larger. To have a better idea of this effect we plotted in
fig. \ref{fig:rat} the relative difference between CRPA and FSI total
cross sections. This figure shows that the relative reduction is of
about the 35\% for neutrinos of 20 MeV and it reaches a value of about
the 10\% above the 40 MeV.

The relevance of FSI in the quasi-elastic excitation region has been
pointed out in other publications \cite{co02,ble01}.  Here we have
shown that for neutrino of few tens of MeV the effects of FSI can be
even more relevant. The consequences on the detection of supernovae
neutrinos are under investigation.

%

%
%
%
%
%
\end{document}